\documentclass[copyright]{eptcs}
\providecommand{\event}{T. Neary, D. Woods, A.K. Seda and N. Murphy (Eds.):\\ The Complexity of Simple Programs 2008.}
\providecommand{\volume}{1}
\providecommand{\anno}{2009}
\providecommand{\firstpage}{153}
\usepackage{breakurl}
\usepackage{graphicx}
\def\ligne#1{\hbox to \hsize{#1}}
\def\PlacerEn#1 #2 #3 {\rlap{\kern#1\raise#2\hbox{#3}}}
\newtheorem{thm}{Theorem}
\newtheorem{lem}{Lemma}

\newtheorem{cor}{Corollary}
\title{On the injectivity of the global function of a cellular automaton in the
hyperbolic plane (extended abstract)}
\author{Maurice Margenstern
   	\institute{Laboratoire d'Informatique Th\'eorique et Appliqu\'ee, EA 3097,\\
        Universit\'e Paul Verlaine $-$ Metz, D\'epartement d'Informatique,\\
        \^Ile du Saulcy, 57045 Metz Cedex, France}
   	\email{margens@univ-metz.fr}
}
\def\titlerunning{On the injectivity of the global function of a cellular automaton in the hyperbolic plane}
\def\authorrunning{M. Margenstern}

\begin{document}
\maketitle

	\begin{abstract}
		In this paper, we look at the following question. We consider cellular automata in
		the hyperbolic plane, see \cite{mmJUCSii,mmkmTCS,mmhedlund,mmbook1} and we 
		consider
		the global function defined on all possible configurations. Is the injectivity
		of this function undecidable? The problem was answered positively in the case
		of the Euclidean plane by Jarkko Kari, in 1994, see \cite{jkari94}. In the present paper,
		we show that the answer is also positive for the hyperbolic plane: the problem is
		undecidable.
	\end{abstract}

\def\cqfd{\hbox{\kern 2pt\vrule height 6pt depth 2pt width 8pt\kern 1pt}}
\def\Hii{$I\!\!H^2$}
\def\Hiii{$I\!\!H^3$}
\def\Hiv{$I\!\!H^4$}
\def\norm{\hbox{$\vert\vert$}}
\section{{\Large Introduction}}

   The global function of a cellular automaton~$A$ is defined in the set of 
all configurations. Note that when we implement an algorithm to solve a given problem,
the initial configuration is usually finite. The study of the global function
starts from another point of view.
  
   In the case of the Euclidean plane, the definition of the set of configurations
is very easy: it is $Q^{Z\!\!Z^2}$, where $Q$~is the set of states of the automaton.

   In the hyperbolic plane, see~\cite{mmbook1,mmhedlund}, we have the following 
situation: we consider that the grid is the pentagrid or the ternary heptagrid, 
see~\cite{mmbook1}. We fix a tile, which will be called the {\bf central cell} 
and, around it, we dispatch $\alpha$~sectors, $\alpha\in\{5,7\}$: $\alpha=5$ in 
the case of the pentagrid, $\alpha=7$ in the case of the ternary heptagrid. We 
assume that the sectors and the central cell cover the plane 
and the sectors do not overlap, neither the central cell, nor other sectors: 
call them the {\bf basic sectors}. Denote by ${\cal F}_\alpha$ the set 
constituted by the central cell and $\alpha$~Fibonacci trees, 
see~\cite{mmJUCSii,mmbook1,mmkmTCS} , each one spanning a basic sector. Then, 
a configuration of a cellular automaton~$A$ in the hyperbolic plane can be 
represented as an element of~$Q^{{\cal F}_\alpha}$, where $Q$~is the set of 
states of~$A$. If $f_A$~denotes the {\bf local} transition function of~$A$, 
its {\bf global} transition function~$G_A$ is defined by: $G_A(c)(x)=f(c(x))$, 
where~$c$ runs over~$Q^{{\cal F}_\alpha}$ and $x\in {\cal F}_\alpha$.

The injectivity problem for a cellular automaton consists in asking whether there
is an algorithm which, applied to a description of~$f_A$ would indicate whether 
$G_A$~is injective or not.

   The goal of this paper is to prove the following result:

\begin{thm}\label{injundec}
There is no algorithm to decide whether the global transition function of a cellular
automaton on the ternary heptagrid is injective or not.
\end{thm} 

   It gives a negative answer to the previous question. Although this answer is
not surprising, it is not at all obvious to prove it. As in the Euclidean case,
the proof relies on the undecidability of the tiling problem. This is a well 
known result for the Euclidean plane, established by R. Berger in 1966, 
see~\cite{berger}, this proof being much simplified by R. Robinson in
1971, see~\cite{robinson1}. The same result for the hyperbolic plane, this 
question was raised by R. Robinson in his 1971 paper was recently only 
established by the present author and J. Kari, in 2007, independently of each 
other, the proofs being completely different.

   In this paper, we give a short account of the proof of the undecidability of
the tiling problem, following the proof I established 
in~\cite{mmBEATCS,mmTCS,mmarXivprefill,mmarXive}, in a new setting I indicated
in~\cite{mmbook2,mmLondon}. This will be Section~2 of this paper.
Section~3 will introduce the mauve triangles, a new object on which
the path, defined in Section~4, can be constructed, yielding the proof
given in Section~5. We shall shortly discuss the situation opened by this
result in our conclusion.

    We shall not repeat the construction of the interwoven triangles
on which the construction of the mauve triangles relies. In section~2, 
we remind the basic properties of the mauve triangles and we introduce 
new ones. In section~3, we more carefully describe the construction of the 
path based on the mauve triangles. In section~4, we show how to derive the 
proof of theorem~\ref{injundec}.

\section{The undecidability of the tiling problem}

   In this section, we just introduce the new setting and then, sketchily 
remember the guidelines of the proof which is explained 
in~\cite{mmBEATCS,mmTCS,mmbook2}.

%\vskip 10pt
%\setbox110=\hbox{\epsfig{file=seeds_new.ps,width=180pt}}
%\setbox112=\hbox{\epsfig{file=seeds_level_new.ps,width=180pt}}
%\vtop{
%\ligne{\hfill
%\PlacerEn {-355pt} {0pt} \box110
%\PlacerEn {-175pt} {0pt} \box112
%}
%\vskip-45pt
\begin{figure}
    \begin{center}
    \includegraphics[width=180pt]{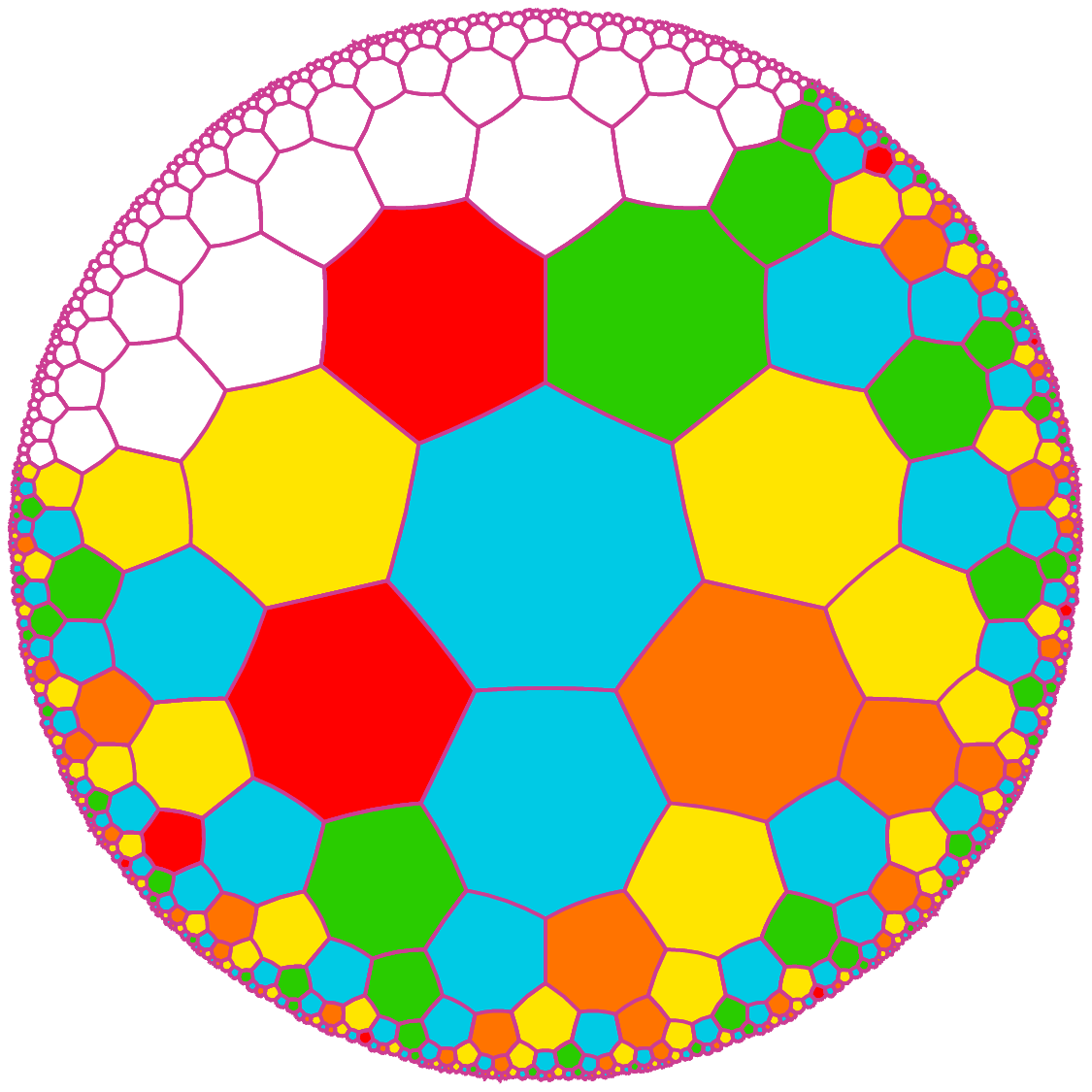}
    \includegraphics[width=180pt]{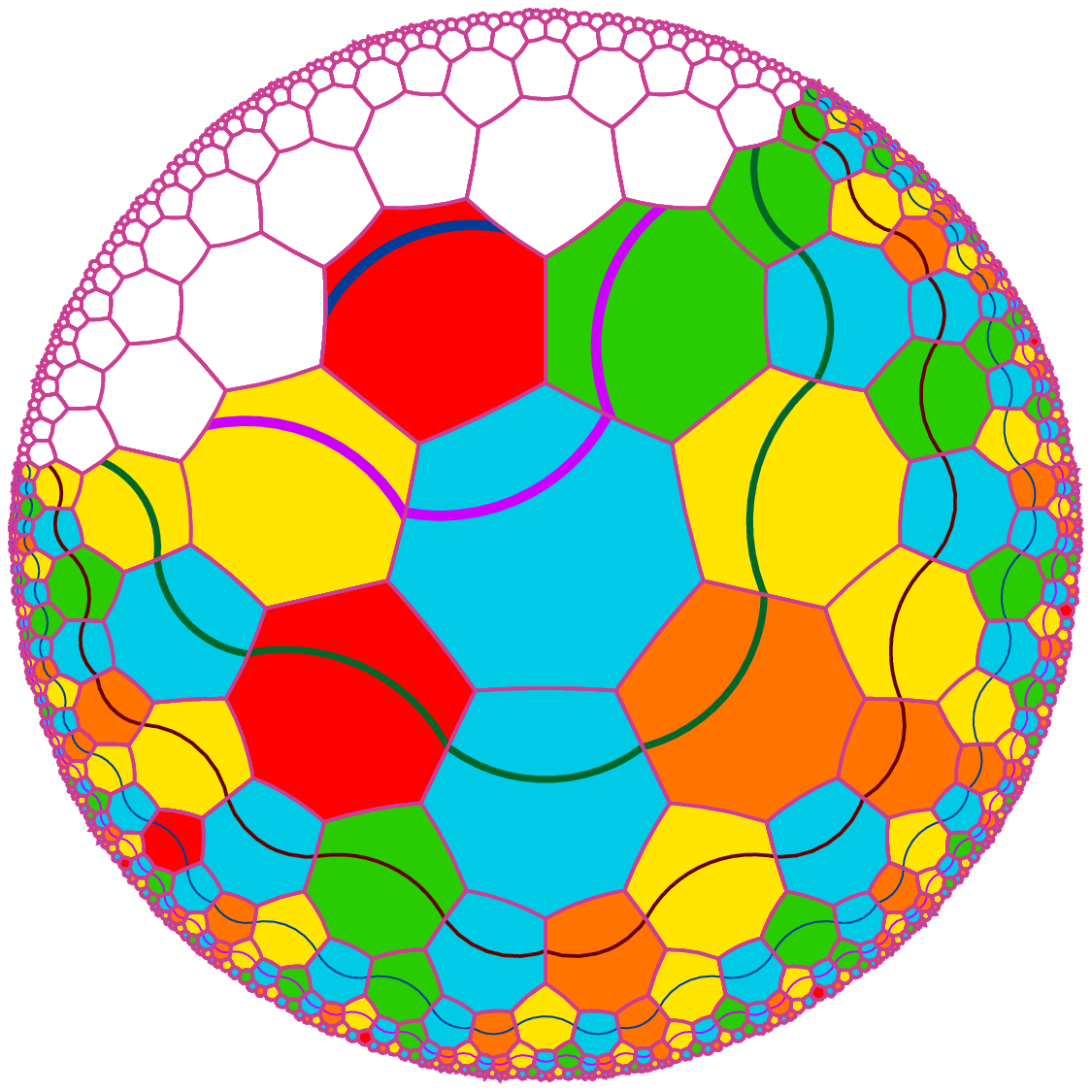}
    \end{center}
\caption{The illustration of the new setting of the construction leading to the proof of 
the undecidability of the tiling problem in the hyperbolic plane.}
    \label{new_sol}
\end{figure}

%\vskip 10pt

  The new setting also takes place on the ternary heptagrid, the tiling $\{7,3\}$
of the hyperbolic plane: it is based on the replication of a regular heptagon
with $\displaystyle{{2\pi}\over3}$ as the interior angle by reflection in its
sides and then, recursively, by reflection of the images in their sides. 

   We introduce four colours for the tiles: green, yellow, blue and orange which
will be denoted by~$G$, $Y$, $B$ and~$O$ respectively. We attach the following
substitution rules, see~\cite{mmbook2,mmLondon}, as illustrated by the left-hand
side picture of Figure~\ref{new_sol}:
\[ G \rightarrow YBG,\ Y \rightarrow YBG,\ B \rightarrow BO,\ O \rightarrow YBO.\]
   As noted in~\cite{mmLondon}, this allows to define two %three 
types of Fibonacci trees, see~\cite{mmJUCSii,mmbook1} for the definition. If we 
define $B$~to be a black node and $G$, $Y$ and~$O$ to be white nodes, we obtain a central
Fibonacci tree, see~\cite{mmJUCSii,mmbook1}. 
%If we define $Y$ and its $G$-son to be black
%and $B$ and~$O$ together with the $G$-son of~$G$ to be white, we obtain a standard 
%Fibonacci tree. At last,
If we define $G$ and~$Y$ to be black and $B$ with~$O$ to be white, then we have
standard Fibonacci trees. %in the display used by the trees of the mantilla,
We define {\bf seeds} as $G$-nodes on an even isocline whose fatheris a $Y$-node
and whose grandfather is a $G$-node.
In Figure~\ref{new_sol}, the tree rooted at a seed, in red in the
figure, implements the representation of the trees of the mantilla,
see~\cite{mmBEATCS,mmTCS,mmbook2}. We use this latter interpretation which allows
us to construct the isoclines which are the horizontals of our constructions, both
for the undecidability of the tiling problem and for the proof of 
Theorem~\ref{injundec}. This is illustrated by the right-hand side picture
of Figure~\ref{new_sol}. The convex arcs are attached to the black nodes,
$G$ and~$Y$, and the concave ones are attached to the white nodes, here $B$ 
and~$O$.

\begin{figure}
    \begin{center}
    \includegraphics[width=400pt]{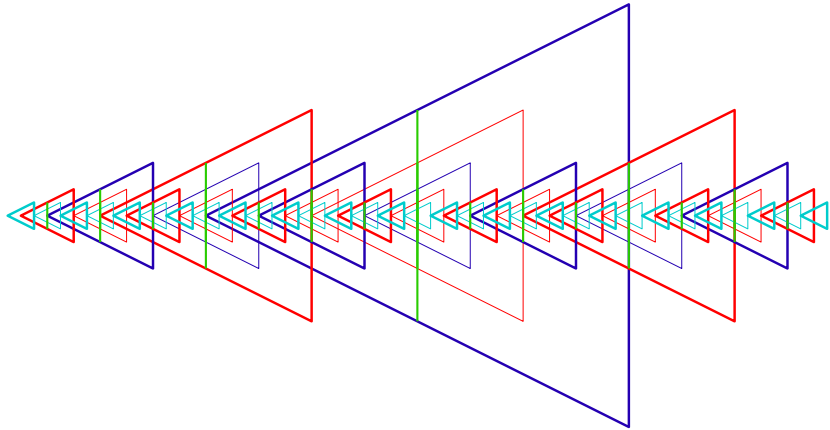}
    \end{center}
\caption{ Construction of the interwoven triangles in the Euclidean plane: look at the green signal.}
    \label{interwovengreen}
\end{figure}

%\vskip -10pt
%\setbox110=\hbox{\epsfig{file=new_interwoven_a5.ps,width=400pt}}
%\vtop{
%\ligne{\hfill
%\PlacerEn {-380pt} {0pt} \box110
%}
%\vskip-45pt
%\begin{fig}\label{interwovengreen}
%\leurre
%\end{fig}
%}

%\vskip 10pt
In this new setting, the isoclines are periodically numbered from~0 
to~3.
   We conclude this section by reminding the principle of the interwoven 
triangles, which is illustrated by Figure~\ref{interwovengreen}. The generation~0
is in light blue in the figure. The even generations are in dark blue. Triangles,
thick strokes, alternate with phantoms, thin strokes, for each generation. 
Triangles and only them generate the vertices and the bases of the
triangles and phantoms of the next generation: light and dark blue triangles 
generate the red generations. The red triangles generate the dark blue 
generations. The reader is referred 
to~\cite{mmBEATCS,mmTCS,mmarXivprefill,mmarXive} for more details on this 
construction and its implementations in the hyperbolic plane.

\section{The mauve triangles}

   The mauve triangles are constructed upon the interwoven triangles.

%, using another tiling as a 
%background. This tiling, called the {\bf mantilla}, is a refinement of the 
%ternary heptagrid by grouping its tile
%in a particular way. Now, it is possible to implement the interwoven triangles 
%in a simpler context of the ternary heptagrid. However, the spacing imposed 
%by the mantilla is a good point which allows to more easily solve a few details 
%of the implementation of the path. This is why, in this paper, we assume the 
%construction to be performed on the mantilla.

   The construction of the interwoven triangles needs a lot of signals, which 
entails a huge number of tiles, still around 9,000 of them in this new context, 
not taking into account the specific tiles devoted to the simulation of a 
Turing machine. The construction of this paper requires much more tiles, but we 
shall not try to count them.

%\subsection{The mauve triangles: properties and constructions}

   The mauve triangles are constructed from the red triangles of 
the interwoven ones. Any mauve triangle is triggered by a red triangle and 
conversely. The vertex of a mauve triangle~$T$ is that of a red triangle~$R$. 
Its legs follow those of~$R$. After the corner of~$R$, they go on, on the same ray, 
until they meet the basis of the red phantoms which are generated by 
the basis of~$R$. At this meeting, the legs meet the basis of the mauve 
triangle which coincide with the basis of the just mentioned red phantoms. 
In~\cite{mmarXivprefill}, we thoroughly describe that this 
construction can be forced by a finitely generated tiling. We refer the reader 
to these papers. 

\subsection{Intersection properties of the mauve triangles}

   As it stems from a red triangle, we say that a mauve triangle of the 
generation~$n$ is constructed on a red triangle of the generation~$2n$+1. Later, 
it will be useful to recognize the mauve triangles of generation~0: they
have the colour {\bf mauve-0} and will be called {\bf mauve-0 triangles}.

   From the doubling of the height with respect to the red triangles, the 
mauve triangles lose the nice property that the red triangles are either 
embedded or disjoint. This is no more the case for the mauve triangles. 
However, the overlappings and intersections of mauve triangles can precisely 
be described.

\begin{figure}
    \begin{center}
    \includegraphics[width=250pt]{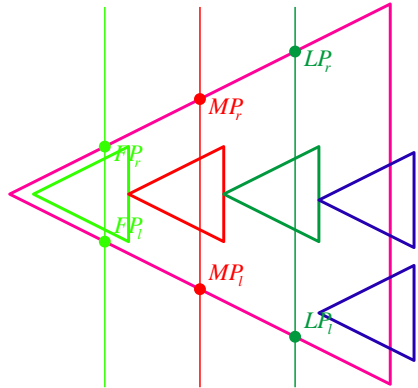}
    \end{center}
\caption{ An illustration of the mauve triangles. Inside the big triangle, from left to
right, the $0$-, $1$-, $2$- and $3$-triangles of the previous generation.} 
    \label{les_mauves}
\end{figure}

%\vskip 5pt
%\vskip 5pt
%\vtop{
%\setbox110=\hbox{\epsfig{file=structure_mauves.ps,width=250pt}}
%%\setbox112=\hbox{\epsfig{file=new_setting_alaCGS.ps,width=110pt}}
%\ligne{\hfill
%\PlacerEn {-135pt} {0pt} \box110
%%\PlacerEn {20pt} {0pt} \box112
%\hfill}
%%\vspace{-90pt}
%\begin{fig}\label{les_mauves}
%\leurre
%An illustration of the mauve triangles. Inside the big triangle, from left to
%right, the $0$-, $1$-, $2$- and $3$-triangles of the previous generation. 
%\end{fig}
%}
%\vskip 5pt
   The figure indicates the {\bf first points}, $FP$, the {\bf mid-points}, $MP$,
and the {\bf low points}, $LP$, of a mauve triangle~$T$ of generation~$n$+1. On 
Figure~\ref{les_mauves}, we can see that the 3-triangles cut the basis of~$T$ 
at their $LP$'s and that the  $LP$'s of the 2-triangles are aligned with those 
of~$T$. The isoclines which pass through the $FP$'s, $MP$'s and $LP$'s of~$T$
cut smaller triangles at their~$LP$'s and for triangles of generation~$n$$-$1
if any, they are cut by these lines if and only if they are 2-triangles. The 
3-triangles of the generation~$n$ which contains the vertex of~$T$, if any,
is called the {\bf hat} of~$T$. Note that inside~$T$, its 0-triangles are all
hatted.

   We can see that the intersection between mauve triangles are those which
are indicated by Figure~\ref{les_mauves} and only those. A mauve triangle~$T$
intersects 3-triangles of the previous generation: always one of them at its 
basis, possibly another one near its vertex, at its legs. Also, $T$~may be
cut by the basis of a mauve triangle of a bigger generation, always at its 
$LP$'s. We refer the reader to~\cite{mmarXivinj} for a proof of these properties.

\subsection{New notions: the $\beta$-clines and the $\beta$-, $\gamma$-points}

   From the fact that the 3-triangles $T_i$'s of the previous generation 
which are inside a mauve triangle~$T$ cut $T$ along its basis and at their $LP$'s,
we can see that we can repeat this observation with the 3-triangles of the 
$T_i$'s until we reach mauve-0 triangles. We call {\bf $\beta$-cline} the isocline
of the mauve-0 triangles reached by this process and we say that the
{\bf $\beta$-cline} is attached to~$T$. Note that the same isocline is the
$\beta$-cline of all the 3-triangles which occur in the just considered 
construction. By definition, we call {\bf $\beta$-points} of~$T$ the intersection
of the legs of~$T$ with the $\beta$-cline of the 2-triangles of the previous
generation which it contains. Similarly, we call {\bf $\gamma$-points} of~$T$
the intersection of the legs of~$T$ with the $\beta$-cline of their hat, if any.
Note that if the hat of~$T$ does not exist, there are infinitely copies of~$T$
within the same set of isoclines which cross~$T$ which are hatted, so that
for these mauve triangles the $\beta$-cline exist and the considered isocline,
the same for all these copies of~$T$ also intersect~$T$, which allows to define
the $\gamma$-points of~$T$.

   In fact, the $\gamma$-points of~$T$ can be constructed in another way, thanks
to the following remark. Assume that the hat~$H$ of~$T$ exists and assume that
$T$~belongs to the generation~$n$+1. Consider a 0-triangle~$Z$ of the 
generation~$n$ which is inside~$T$. It is plain that the hat of~$Z$ exists
if $n>0$, as it is inside~$T$ and that it is a 3-triangle of the 
generation~$n$$-$1 which cuts the basis of~$H$. Accordingly, $H$ and the hat 
of~$Z$ have the same $\beta$-cline. Consequently, to find the $\gamma$-point 
of~$T$, it is enough to take the intersection of the legs of~$T$ with the 
$\beta$-cline of the hat of~$Z$. Note that if~$Z$ is a mauve-0 triangle, the considered 
$\beta$-cline is the isocline which passes through the vertex of~$Z$.
Consequently, this defines the $\gamma$-points of~$T$, independently from the 
existence or not of the hat of~$T$.

Note that the $\beta$-, and $\gamma$-points are not defined for mauve-0 triangles.
\vskip 5pt
   All the points which we defined in Subsection~3.1 and in this one can be 
constructed by a small set of signals in terms of tiles. 

   The constructions relies on the fact that the isoclines joining the $FP$'s and
the $MP$'s as well as the basis of~$T$ have the following property, assuming
that $T$ belongs to the generation~$n$+1: if the isocline meets a triangle
of the generation~$i$ with $i<n$, it is a 2-triangle. This allows to locate
the 0-, 1- and 3-triangles of the generation~$n$ whose vertex is inside~$T$.
From this detection, we also can detect the 2-triangles of the generation~$n$
which are inside~$T$.

   Now, we have to indicate how to construct the $LP$'s of~$T$ as well as 
its $\beta$-cline.

   The construction of the $LP$'s of~$T$ start from the $MP$'s which send
two signals: one along the leg of~$T$ and another along the isocline joining
the $MP$'s, in the direction of the other~$MP$. When the horizontal signal
meets the vertex of a red phantom, it goes down along it, until it reaches
the mid-point~$M$ of the leg of the phantom. There, it goes back to the leg of~$T$
by following the isocline of~$M$, circumventing all triangles which are met from
their right-, left-hand side leg, the signal emanating from the left-, 
right-hand side leg of~$T$. In this way, the first leg met by the signal whose 
laterality is that of the leg from which it came is the expected leg and it also
meets the second signal sent by the~$MP$.

   The construction of the $\beta$-cline of~$T$ follows from this construction.
A signal leaves each $LP$ of~$T$. They go to the opposite side. They follow the
leg of~$T$, then its basis and then, each time they meet a leg of a 3-triangle
at its $LP$, they go down along this leg and, finding the corner, they go along 
the isocline of the basis of the corner, in the direction of the other corner.
This signal is permitted only to go down or to follow an isocline. Consequently,
the two considered signals will meet on the isocline which is the $\beta$-cline
of~$T$, which provides the construction.

\begin{figure}
    \begin{center}
    \includegraphics[width=0.45\textwidth]{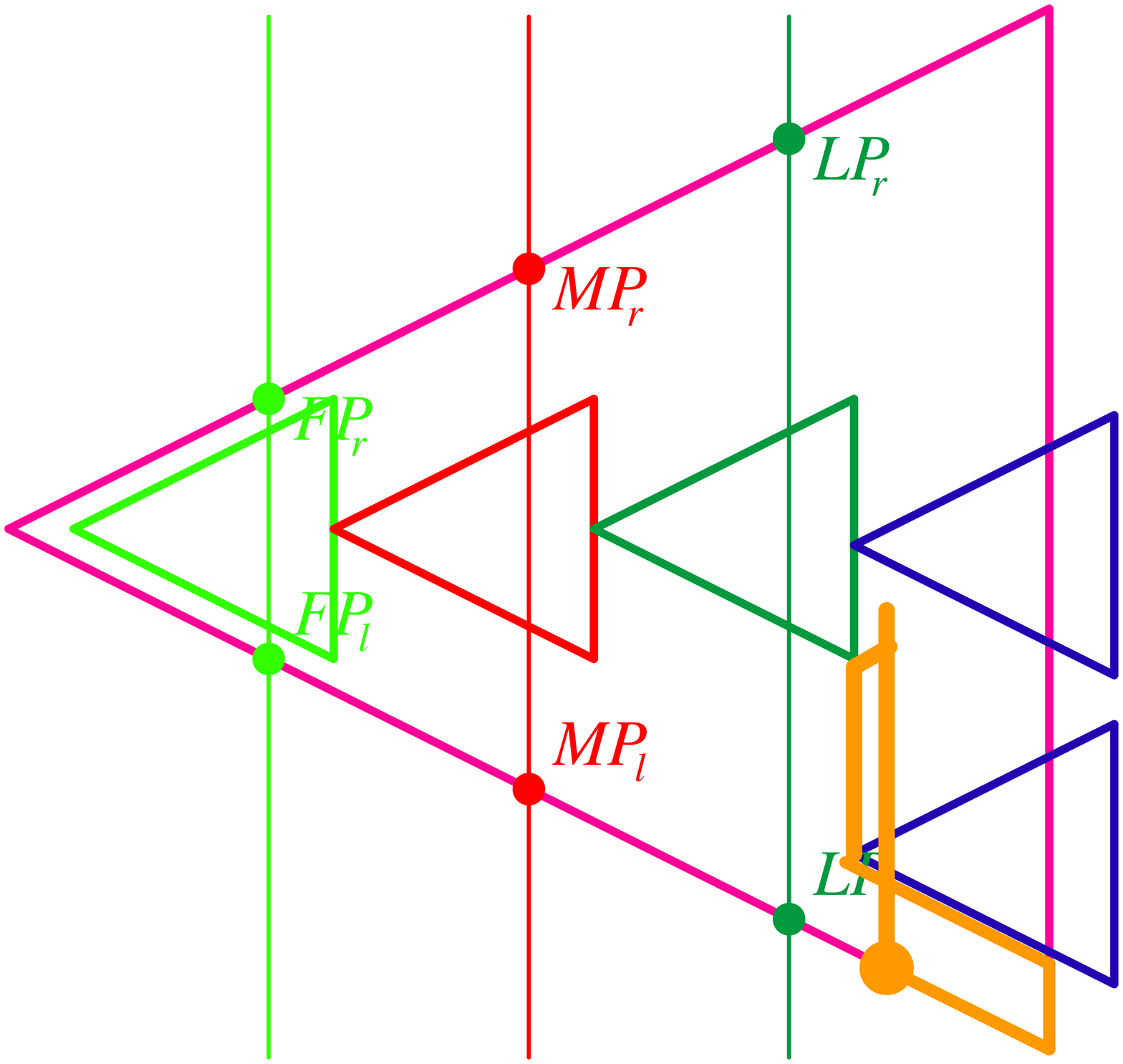}
    \includegraphics[width=0.45\textwidth]{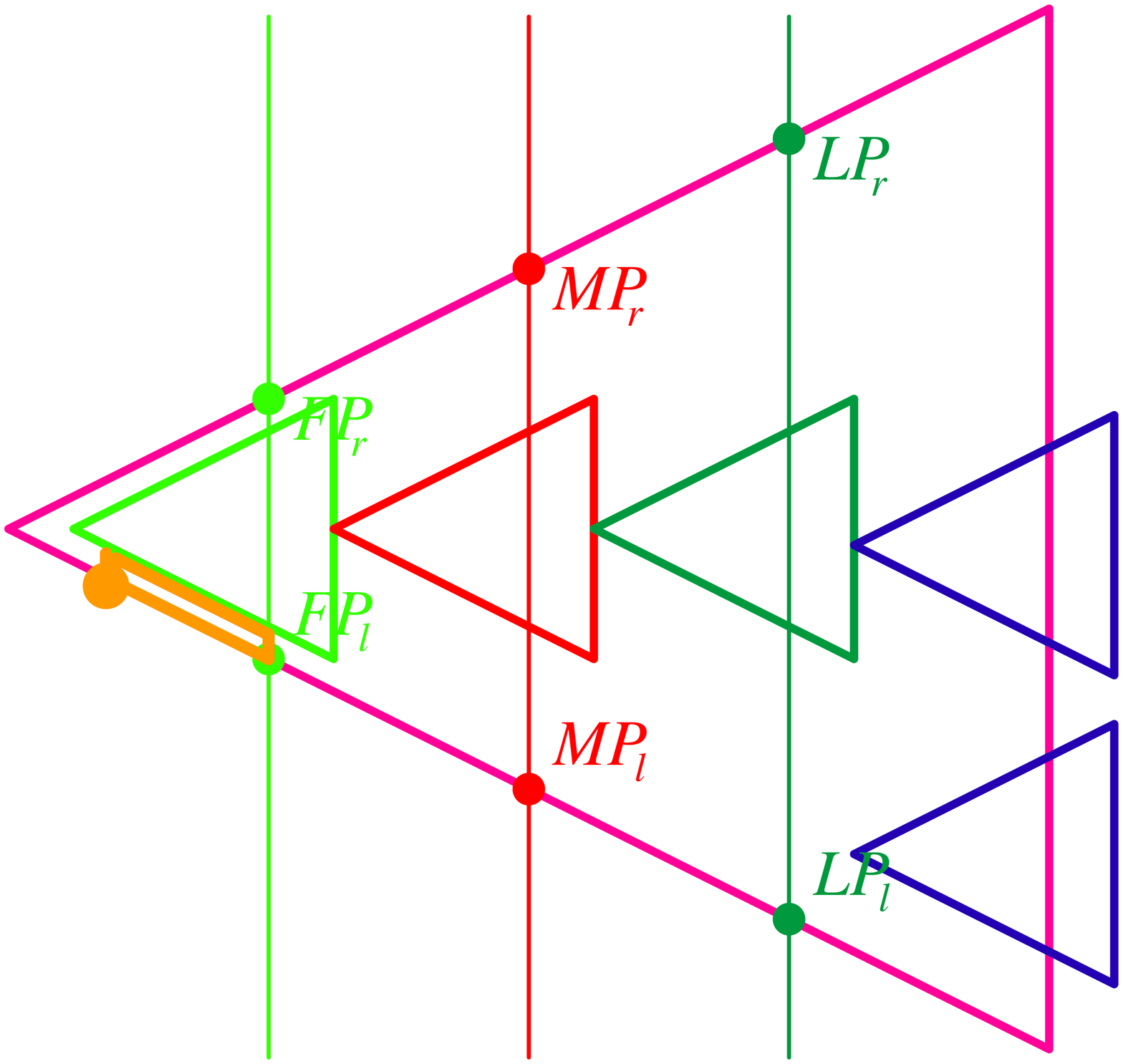}
    \end{center}
\caption{ Illustration of the construction of the $\beta$- and $\gamma$-points of
a mauve triangle of a positive generation.
On the left-hand side: construction of the $\beta$-point. On the right-hand side,
construction of the $\gamma$-point.}
    \label{betagamma}
\end{figure}

   The detection of a $\beta$-cline can now be used to construct the $\beta$-
and $\gamma$-points, combining the detection of a $\beta$-cline with that of
a 0-, or a 2-triangle. The pictures of Figure~\ref{betagamma} illustrates
the constructions of the $\beta$- and the $\gamma$-points of~$T$.

%\vskip 5pt
%\vskip 5pt
%\vtop{
%\setbox110=\hbox{\epsfig{file=beta_construction.ps,width=250pt}}
%\setbox112=\hbox{\epsfig{file=gamma_construction.ps,width=250pt}}
%%\setbox112=\hbox{\epsfig{file=new_setting_alaCGS.ps,width=110pt}}
%\ligne{\hfill
%\PlacerEn {-215pt} {0pt} \box110
%\PlacerEn {-25pt} {0pt} \box112
%%\PlacerEn {20pt} {0pt} \box112
%\hfill}
%%\vspace{-90pt}
%\begin{fig}\label{betagamma}
%\leurre
%Illustration of the construction of the $\beta$- and $\gamma$-points of
%a mauve triangle of a positive generation. 
%\vskip 0pt
%On the left-hand side: construction of the $\beta$-point. On the right-hand side,
%construction of the $\gamma$-point.
%\end{fig}
%}
%\vskip 5pt
%The $HP$'s also are not difficult to be determined: they are the intersection of the 
%leg of~$T$ with the basis of its hat.
%Note, that the $HP$ does not exist for the mauve triangles of generation~0. 
%When $T$~blongs to a positive generation, the $HP$ is the first point of the 
%segment of the leg from the mid-point to the vertex where a basis is encountered.
In \cite{mmarXivinj}, we completely prove the following statement:
  
\begin{lem}\label{tileLP}
The mauve triangles together with the determination of their $LP$'s and 
mid-points can be constructed from a finite set of prototiles.
\end{lem}
\vskip 5pt
\noindent
which can be established using the above ideas.

   The $\gamma$-point allows us to define a last couple of points on the legs of
a mauve triangles, the $HP$'s, {\it i.e.} {\bf high points}.

   The $HP$'s are defined with respect to the $\gamma$-points as follows. If the
$\beta$-cline which crosses the $\gamma$-point is a $\beta$-cline of type~2, 
then the $HP$ is the $\gamma$-point. If this is not the case, then the $HP$ of~$T$
is defined by the first intersection with a basis of a mauve triangle with the  
legs of~$T$, when starting from the $FP$'s and going up towards the vertex.
If no basis is met before reaching the vertex, the $HP$ of~$T$ is the intersection
of the legs with the isocline which is just below that of the vertex of~$T$. This
is the case for the mauve-0 triangles.

%Secondly, the yellow 
%signal cannot cross legs. Thirdly, it cannot meet a leg from inside: always from outside. 
%In other words, the yellow signal cannot join two legs of the same triangle on the same 
%isocline. But it can do that for two legs of opposite lateralities of consecutive 
%triangles of the same positive generation within the same latitude. 

%   The condition of no meeting of a leg from inside means that the join tile exists in
%three versions. In one version, the yellow signals goes from one side of the tile to the
%other. In a second version, it goes from the centre of the tile to one side and, on the
%third verion, it is the symmetrical configuration involving the other side. This allows
%us to use the correct version
   
%%  il ne faut pas supprimer le signal jaune : il faut le doubler par un noir
%%  si le $\beta$-signal descendant est présent

\section{A half-plane filling path}

   Now, we turn to the construction of the path.

   As we shall see, the path has no cycle and most often, it consists of a single
component which is plane-filling path. In one exceptional situation, the path
breaks down into infinitely many infinite components. There is no cycle and 
each component fills up at least a half-plane.

   The points which we defined in the previous section can be seen as signboards
which are placed on the path in order to indicate it in which regions it passes
and which direction it has to follow.

   Our first task consists in describing these regions. In a second step, we shall
see how the path fills up the regions, on the basis of an inductive construction.
%% panneau de signalisation : road sign, signboard
%% panneau indicateur : signboard

   We define two types of basic regions. The first one is the set of tiles 
defined by a mauve triangle: its borders and its inside. Remember that the 
basis of a mauve triangle contains more than the majority of tiles resulting 
from the just given definition. It is considered as a basic region as once 
the path enters a mauve triangle~$T$, it fills up~$T$ almost completely before 
leaving~$T$. In fact, there is a restriction and the path fills a bigger area. 
Remember that the basis of~$T$ is crossed by the legs of a lot of triangles,
the 3-triangles of the previous generation whose vertices are inside~$T$ and
a lot of 2-triangles of smaller generations. The path enters each such 
triangle~$M$ through one of its $LP$'s and it fills it up completely, leaving
it through the $HP$~of~$M$ which lies on the other leg of~$M$. After a while,
as will be seen, the path goes back to the basis of~$T$ and will meet the
next small triangle which crosses this basis. Of course, inside a small triangle
filled up by the path, the process is recursively repeated: if the basis of~$M$
is crossed by legs of smaller triangles, the path fills them up before going
on on the basis of~$M$.

   This process allows us to define the latitude of a mauve triangle~$T$.
For a mauve-0 triangle, the lower border is the isocline of the basis of~$T$
and it belongs to the latitude. The upper border is the isocline of the vertex
of~$T$ and it belongs to another latitude. For higher generations, the border
of the latitude belonging to the latitude is defined by the isocline of the
basis of~$T$ with this exception that each time this isocline is crossed by
legs of smaller triangles, the border goes down along the legs and then follows
the basis of the smaller triangles, repeating this way recursively, until a 
mauve-0 triangle is reached.

\begin{figure}
    \begin{center}
    \includegraphics[width=280pt]{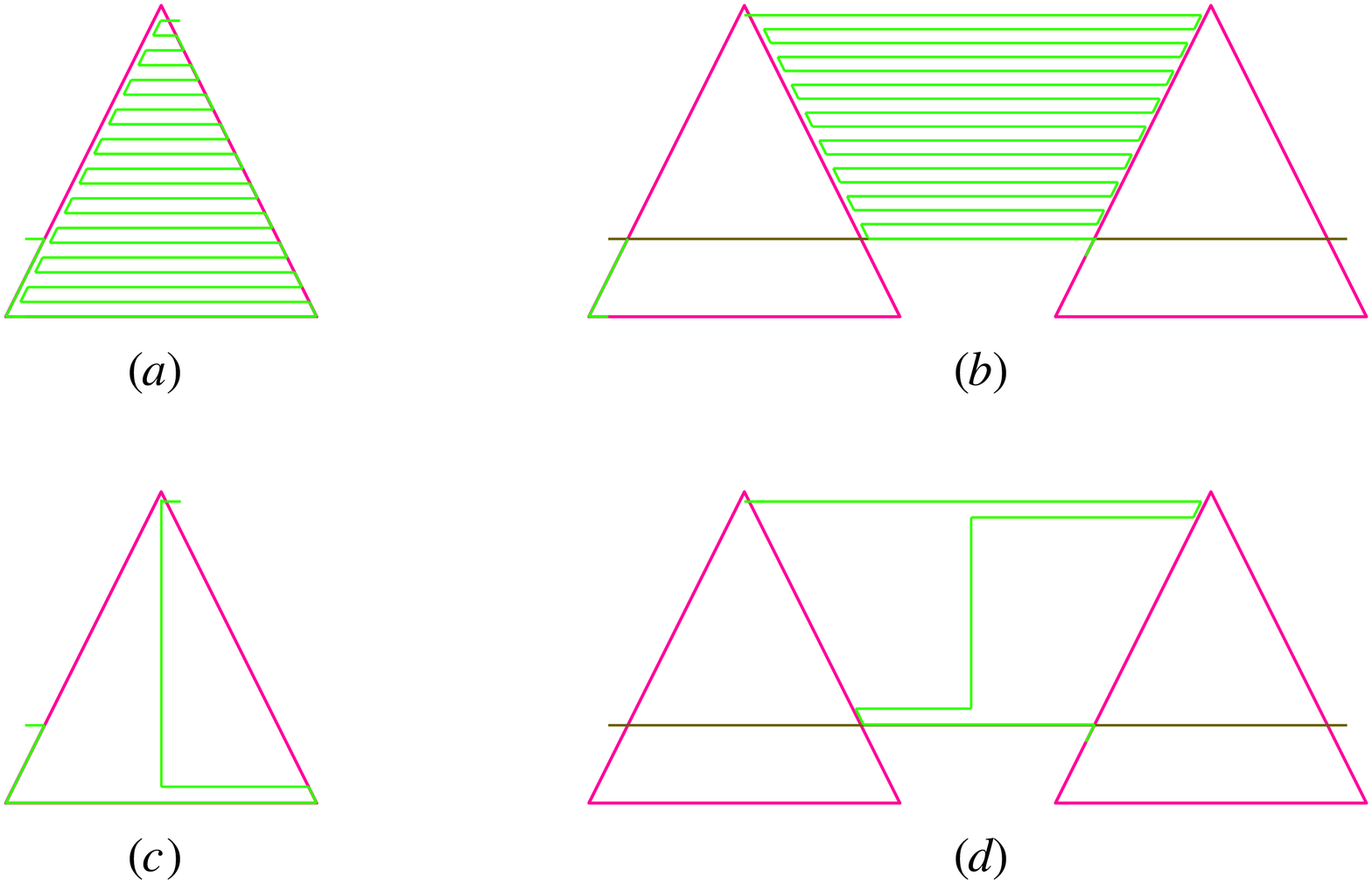}
    \end{center}
\caption{A schematic representation of the path:\protect\\ On the left-hand side, inside a mauve-0 triangle. On the right-hand side, 
in between two consecutive mauve-0 triangles within the same latitude when the
vertices belong to the same basis.}
    \label{schemas}
\end{figure}

\begin{figure}
    \begin{center}
    \includegraphics[width=360pt]{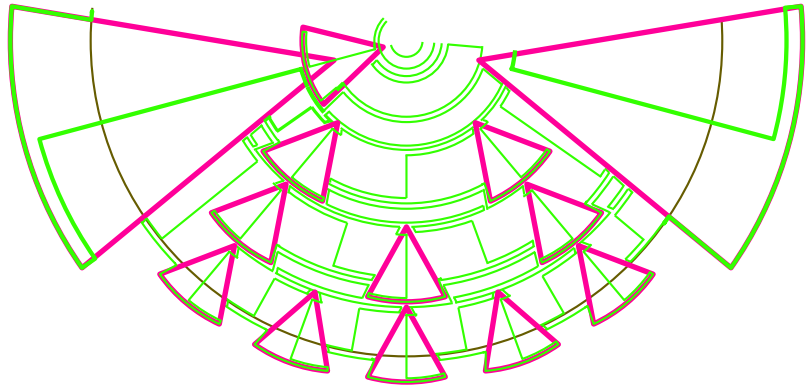}
    \end{center}
\caption{ The path in between two triangles of generation~$1$.}
    \label{mauve_1}
\end{figure}

%\vskip 10pt
%\setbox110=\hbox{\epsfig{file=schemas.ps,width=280pt}}
%%\setbox112=\hbox{\epsfig{file=new_setting_alaCGS.ps,width=110pt}}
%\vtop{
%\ligne{\hfill
%\PlacerEn {-140pt} {0pt} \box110
%%\PlacerEn {20pt} {0pt} \box112
%\hfill}
%\vspace{-5pt}
%\begin{fig}\label{schemas}
%\leurre
%A schematic representation of the path:
%\vskip 0pt
%\noindent
%On the left-hand side, inside a mauve-0 triangle. On the right-hand side, 
%in between two consecutive mauve-0 triangles within the same latitude when the
%vertices belong to the same basis.
%\end{fig}
%}
%\vskip 7pt
   In Figure~\ref{schemas}, the picture~$(a)$ illustrates the path inside
a mauve-0 triangle and the picture~$(b)$ illustrates the path in between two
consecutive mauve-0 triangles inside the same latitude. Now, for this situation
in between two consecutive mauve triangles of the same latitude, the path
fills up the space between the upper and the lower borders. The situation is
in general different from the picture~$(b)$. Figure~\ref{mauve_1} describes
the general situation for consecutive mauve triangles of generation~1 
standing within the same latitude. There, we can see the various configurations
which may occur in between consecutive mauve-0 triangles of the same latitude.

%%\ligne{\hfill}
%\vskip-100pt
%\setbox110=\hbox{\epsfig{file=new_new_new_cmauve_1_inbetween.ps,width=360pt}}
%%\setbox112=\hbox{\epsfig{file=new_setting_alaCGS.ps,width=110pt}}
%\vtop{
%\ligne{\hfill
%\PlacerEn {-160pt} {0pt} \box110
%%\PlacerEn {20pt} {0pt} \box112
%\hfill}
%\vspace{-25pt}
%\begin{fig}\label{mauve_1}
%\leurre
%The path in between two triangles of generation~$1$.
%\end{fig}
%}

\begin{figure}
    \begin{center}
    \includegraphics[width=0.45\textwidth]{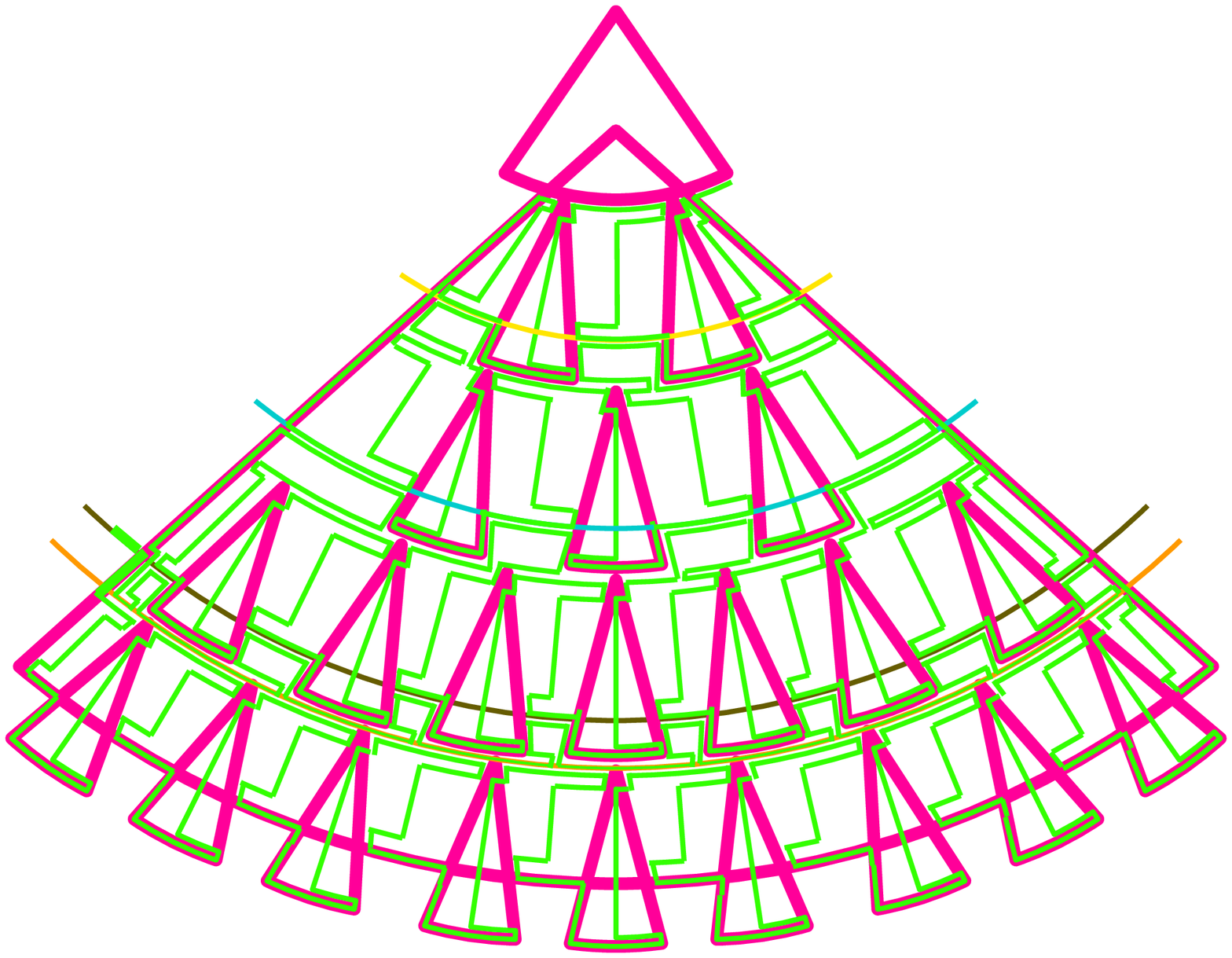}
    \includegraphics[width=0.45\textwidth]{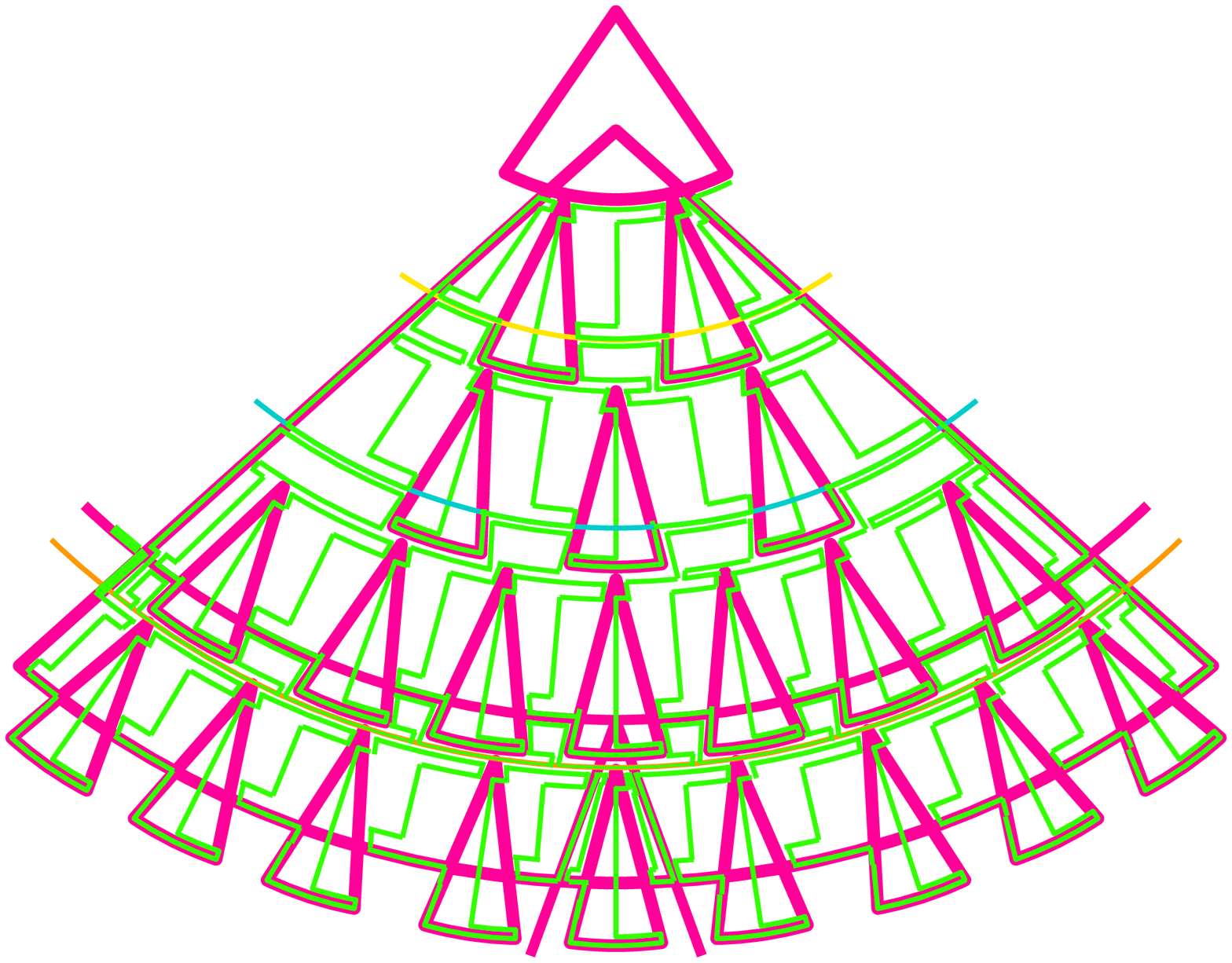}
    \end{center}
\caption{ The path inside a triangle of generation~$1$.\protect\\
On the left-hand side: a $0$- or a $1$-triangle. On the right-hand side, a~$3$-triangle 
or a $2$-triangle which is cut by a mauve triangle of a bigger generation.}
    \label{mauve_1in}
\end{figure}

%\vskip -5pt
   In Figure~\ref{mauve_1} we make use of the schematic representation used
the pictures~$(c)$ and~$(d)$ of Figure~\ref{schemas} to represent the
pictures~$(a)$ and~$(b)$. This is repeated in Figure~\ref{mauve_1in} which
represents the path inside a mauve triangle of generation~1.

%\ligne{\hfill}
%\vskip-105pt
%\setbox110=\hbox{\epsfig{file=cmauve_in_triangle1.ps,width=280pt}}
%\setbox112=\hbox{\epsfig{file=cmauve_in_triangle1_bis.ps,width=280pt}}
%\vtop{
%\ligne{\hfill
%\PlacerEn {-210pt} {0pt} \box110
%\PlacerEn {-40pt} {0pt} \box112
%\hfill}
%%\vspace{-5pt}
%\begin{fig}\label{mauve_1in}
%\leurre
%The path inside a triangle of generation~$1$.
%\vskip 0pt
%On the left-hand side: a $0$- or a $1$-triangle. On the right-hand side, a~$3$-triangle 
%or a $2$-triangle which is cut by a mauve triangle of a bigger generation.
%\end{fig}
%}

   Other details attached to the path are dealt with in~\cite{mmarXivinj}
and we refer the reader to this paper.

   Now, we notice that Figures~\ref{mauve_1} and~\ref{mauve_1in} can also
illustrate the passage from the generation~$n$ to the generation~$n$+1. 
We draw the attention of the reader on the following point. Due to the position
of the entry and the exit points for the path in a triangle, the path fills up
the bottom part of the triangle which is below the isocline joining its $LP$'s
in an almost cycle and then it crosses the other three strips roughly 
corresponding to 2-, 1- and 0-triangles of the previous generation, in changing
its direction each time it enters a new strip. We have a similar feature
in between two consecutive triangles of the same latitude. This dictates a
difference of behaviour when going to a strip to another. Now, this is
regulated by the $\beta$-clines of type~2: they delimit the bottom part of
inside a triangle and they delimit the upper one in between two consecutive
triangles of the same latitude. As by definition, the intersection of 
a $\beta$-cline of type~2 with the legs of a triangle define the $HP$ of this 
triangle, this allows to give a passage to the path for going back to the
origin of the almost cycle. The intersection of a $\beta$-cline which is not of
type~2 with the legs of a triangle prevent the path to enter the triangle or 
to exit from it.

   With these indications, we conclude that the construction is correct and that
the path possesses the following property:

\begin{lem}\label{nocycle}
The path contains no cycle.
\end{lem}

%\noindent
%Proof. I there was a cycle, it would be contained in some basic region. Now, the path
%enters the region through an $LP$ in case of a triangle, through an $HP$ in case of an
%in between region and it exits through the opposite~$HP$ or $LP$ respectively. Accordingly,
%there is no cycle in this region due to the recursive structure of the basic regions
%and it is plain that there is no cycle in a basic region of generation~0. \cqfd

%Another one is also very important:

\begin{lem}\label{anysize}
For any tile~$\tau$, the path on one side of~$\tau$ fills up infinitely many mauve
triangles of increasing sizes.
\end{lem}

%\noindent
%Proof. This also comes from the filling up of the basic regions and their recursive
%structure. The path cannot remain in the same latitude for ever. And so, it goes to another
%latitude, upper or lower, depending on the structure of the underlying model implemented
%by the interwoven triangles. This ensures the conclusion of the corollary. \cqfd

From these lemmas, as established in~\cite{mmarXivinj}, we get:

\begin{cor}\label{onecomponent}
If there are only finite basic regions, the path goes through any tile of the plane.
\end{cor}

   Now, there is an exceptional situation in which we have infinitely
many infinite triangles. In fact, as the path fills up all triangles, the
infinite ones included, it splits into infinitely many components which still
have no cycle and which fill up the region contained in an infinite triangle
and the one in between this infinite triangle and the next one in the same
infinite latitude. Now, for these components, Lemmas~\ref{nocycle} 
and~\ref{anysize} are still true, see \cite{mmarXivinj}.

\section{Proof of the main theorem}

   We can now prove Theorem~\ref{injundec}.
   
   The proof follows the argument of~\cite{jkari94}, with an adaptation
to the case of infinite triangles.

   We define a direction for the path constructed in Section~4. To this aim,
we introduce three hues in the colour used for the signal of the path. One 
colour calls the next one and the last one calls the first one. The periodic 
repetition of this pattern together with the order of the colours define 
the direction. This allows to define the successor of a tile on the path,
which we formalize by a function $\delta$ 
from $Z\!\!\!Z$ to the tiling such that $\delta(n$+$1)$ is the successor of
$\delta(n)$ on the path.

   Consider $M$ a deterministic Turing machine with a single head and a 
single bi-infinite tape which is assumed to be initially empty. 
From~\cite{mmBEATCS,mmTCS}, we can define a finite set of tiles~$T_M$ such 
that $T_M$ tiles the hyperbolic plane if and only if~$M$ does not halt. %As in
%   Remember that in
%~\cite{jkari94}, a
When $T_M$~tiles the hyperbolic plane, it defines infinitely many triangles of infinitely
many sizes and in each triangle, the simulation of~$M$ is resumed from its beginning,
running until the basis of the triangle is reached.
An automaton~$A_M$ is attached to~$M$ and its states are defined by
%a set of tiles~$T$ has its states in 
$D\times\{0,1\}\times T_M$, where $D$ is the set of tiles 
which defines the tiling which we have constructed in Sections~3 and~4. 
The ${0,1}$-component of a state is called its {\bf bit}.
%with the plane filling property and~$T$ is an arbitrary 
%finite set of tiles. 
We can still tile the plane as %we assume that 
the tiles of~$T_M$ are ternary heptagons but the abutting conditions may be 
not observed: if it is observed with all the neighbours of the cell~$x$, the 
corresponding configuration is said to be {\bf correct} at~$x$, otherwise it 
is said {\bf incorrect}. When the considered configuration is correct at every
tile for~$D$ or at every tile for~$T_M$, it is called a {\bf realization} of the 
corresponding tiling. 

As in~\cite{jkari94}, the transition function does not change neither 
the $D$- nor the \hbox{$T_M$-component} of the state of a cell~$x$: it only 
changes its bit. As in~\cite{jkari94}, we define $A_M(c(x))=c(x)$ if the 
configuration in~$D$ or in~$T$ is incorrect at the considered tile. If both are correct, 
we define $A_M(c(x))=\hbox{\rm xor}(c(x),c(\delta(x)))$. It is plain that 
if~$M$ does not halt, $T_M$ tiles the hyperbolic plane and there is a configuration 
of~$D$ and one of~$T_M$ which are realizations of the respective tilings. Then, the 
transition function 
computes the xor of the bit of a cell and its successor on the path. Hence, defining 
all cells with~0 and then all cells with~1 define two configurations which~$A_M$ 
transform to the same image: the configuration where all cells have the bit~0. 
Accordingly, $A_M$ is not injective.

Conversely, if $A_M$~is not injective, we have two different 
configurations~$c_0$ and~$c_1$ for which the image is the same. Hence, there is a 
cell~$x$ at which the configurations differ. Hence, the xor was applied, which means 
that~$D$ and~$T$ are both correct at this cell in these configurations and it is not 
difficult to see that the value for each configuration at the successor of~$x$ on
the path must also be different. And so, following the path in one direction, we have a 
correct tiling for both~$D$ and~$T_M$. Now, from Lemma~\ref{anysize},
as the path fills up infinitely many triangles of increasing sizes, this
means that the tiling realized for $T_M$ is correct in these triangles. 
But, in each triangle, the computation of~$M$ is simulated from its beginning.
Accordingly, the Turing machine~$M$ never halts. And so, we proved that $A_M$ is not 
injective
if and only if $M$ does not halt. Accordingly, the injectivity of $A_M$ is undecidable.
\cqfd

\section{Conclusion}

   In the Euclidean plane, the surjectivity of the global function of cellular 
automata is tightly connected with the injectivity of the function. This is 
not the case in the hyperbolic plane, see~\cite{mmBristol}, so that the 
question of the surjectivity is open in this case.
    Note that the present construction can be generalized to any grid
$\{p,3\}$ of the hyperbolic plane, with $p\geq7$. 

%\section*{Acknowledgment}
%
%I wish to express special thanks to Chaim Goodman-Strauss for his great attention
%to my works and for very encouraging and enlightning comments.
 
\bibliographystyle{eptcs}

\end{document}